\documentclass[aip,rsi,amsmath,amssymb,reprint]{revtex4-1}
\usepackage{graphicx}
\usepackage{dcolumn}
\usepackage{bm}
\usepackage{color}
\newcommand{\ic}{``}

\begin{document}

\preprint{AIP/123-QED}

\title{Design and construction of a point-contact spectroscopy rig with lateral scanning capability}

\author{M. Tortello}\affiliation{Dipartimento di Scienza Applicata e Tecnologia, Politecnico di Torino, Torino 10129, Italy}
\author{W. K. Park}
\thanks{Author to whom correspondence should be addressed. Electronic mail:
wkpark@illinois.edu}
\author{C. O. Ascencio}
\author{P. Saraf}
\author{L. H. Greene}
\thanks{Current address: National High Magnetic Field Laboratory and Department of Physics, Florida
State University, Tallahassee, Florida 32310, USA}
\affiliation{Department of Physics and Materials Research Laboratory, University of Illinois at
Urbana-Champaign, Urbana, Illinois 61801, USA}


\date{\today}

\begin{abstract}
The design and realization of a cryogenic rig for point-contact spectroscopy measurements in the needle-anvil configuration is presented. Thanks to the use of two piezoelectric nano-positioners, the tip can move along the vertical ($z$) and horizontal ($x$) direction and thus the rig is suitable to probe different regions of a sample \textit{in situ}. Moreover, it can also form double point-contacts on different facets of a single crystal for achieving, e.g., an interferometer configuration for phase-sensitive measurements. For the latter purpose, the sample holder can also host a Helmholtz coil for applying a small transverse magnetic field to the junction. A semi-rigid coaxial cable can be easily added for studying the behavior of Josephson junctions under microwave irradiation. The rig can be detached from the probe and thus used with different cryostats. The performance of this new probe has been tested in a Quantum Design PPMS system by conducting point-contact Andreev reflection measurements on Nb thin films over large areas as a function of temperature and magnetic field.

\end{abstract}

\maketitle

\section{\label{sec:1}Introduction}
\begin{figure}[btp]
\begin{center}
\includegraphics[keepaspectratio, width=0.9 \columnwidth]{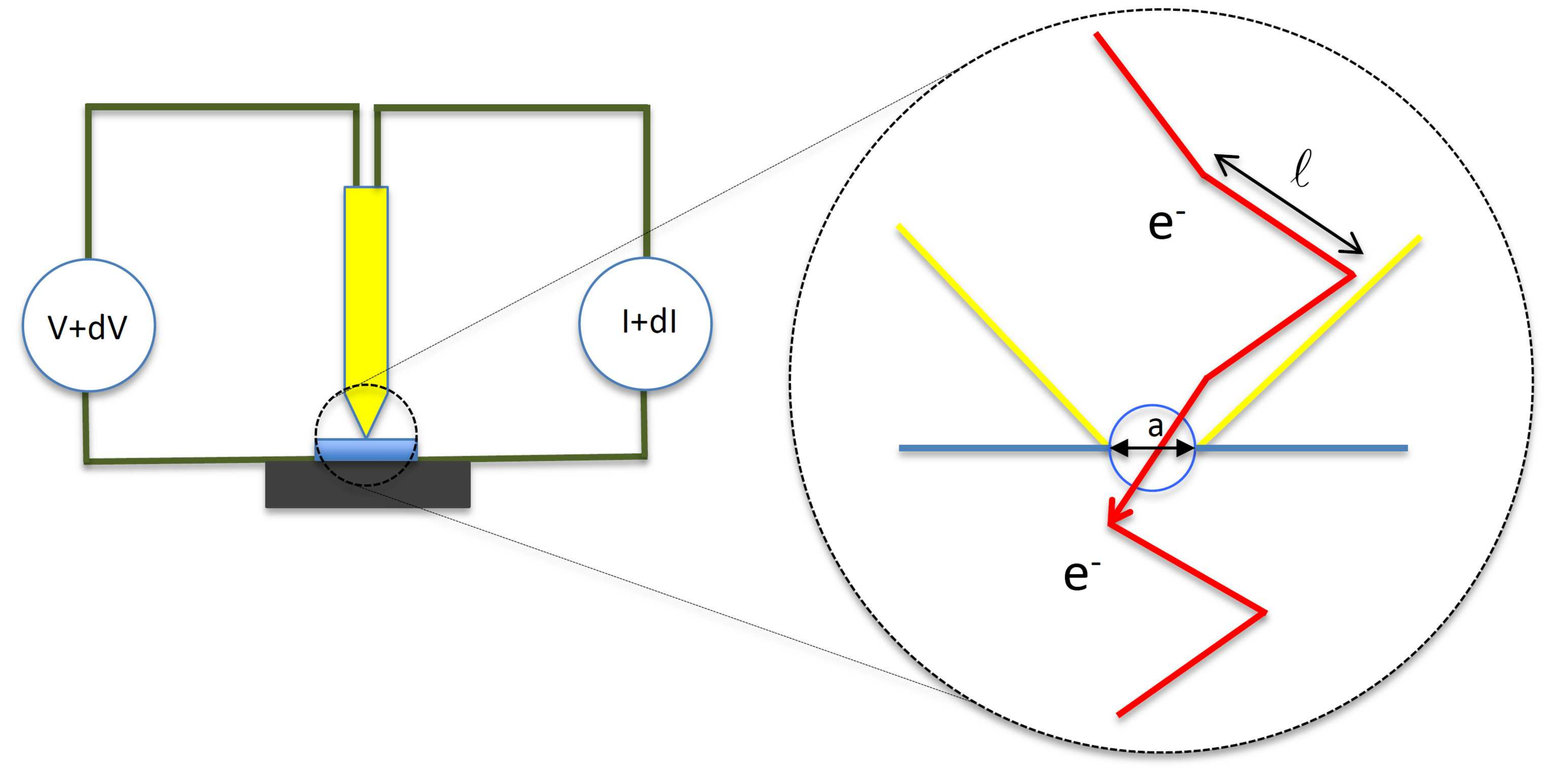}
\end{center}
\vspace{-5pt}
\caption{Experimental arrangement for PCS in the so called needle-anvil configuration. A sharp metallic tip is brought into contact with a sample under study. The dashed circle on the right encloses a schematic of the ballistic flow of electrons with mean free path $\ell$ through the point contact with characteristic size $a$, when $a\ll \ell$.}\label{n-a}
\end{figure}

Point-contact spectroscopy (PCS) was introduced by I. K. Yanson more than 40 years ago,\cite{yanson74} when he noticed some nonlinearities in the current-voltage characteristics (and in its second derivative) of a tunnel junction between two metals in the presence of micro-shorts in the tunnel barrier. It turned out that the observed nonlinearities contained information on the phonon spectrum of the investigated material. Indeed, if the characteristic size $a$ of the micro-constriction or \ic point-contact" is smaller than the electron mean free path $\ell$, the electrons can be injected ballistically, arriving on the other side of the contact with known excess energy, and eventually be inelastically scattered by the phonons of the investigated material. These events cause a negative correction term to the current caused by the electrons back-scattered inelastically losing an energy corresponding to that of the relevant phonons. This mechanism allows obtaining spectroscopic information on the material under study and the electron-phonon spectral function was thus determined for many elemental metals and alloys.\cite{naidyuklibro} It was also subsequently found that this process occurs for other elementary excitations and scattering mechanisms in the materials, like electron-magnon interaction, scattering with magnetic impurities, and so on.\cite{naidyuklibro} Hence comes also the name of quasiparticle scattering spectroscopy (QPS).\cite{park12} There can be different experimental realizations of point contacts \cite{naidyuklibro} but the most popular one is certainly the so-called needle-anvil technique,\cite{jansen77,park06} in which a very sharp metallic tip is gently brought into contact with the sample, as depicted in Fig. \ref{n-a}.

PCS became also very popular for the study of superconducting materials. Indeed, if the sample is a superconductor, phenomena like Andreev reflection \cite{andreev64} or quasiparticle tunneling occur and, thanks to the development of suitable theoretical models for analyzing the data,\cite{BTK,kashiwaya96,daghero10} it became possible to determine the number, amplitude, and symmetry of the superconducting order parameter in several classes of materials.\cite{deutscher05,daghero11,park08,park09b,lu11}
If the tip is also made of a superconducting material, a Josephson junction of the weak link type \cite{likharev79} can be formed. Although in view of superconductive electronics applications point-contact Josephson junctions lack the desired reproducibility, they nevertheless represent one of the easiest ways to form a Josephson junction \cite{golubov04} and can be suitably used to investigate the fundamental properties of novel superconductors.\cite{zhang09,burmistrova15} It is also worth recalling here that double point-contact junctions in an interferometer configuration have been used \cite{brawner94} or proposed \cite{golubov13} in order to clarify the order parameter symmetry of unconventional superconductors. Finally, this laboratory has recently shown that PCS can be used to identify and filter out density of states arising from strong correlations.\cite{park12,park14,arham12,lee12,lee13,lee15,leegreene15}

PCS probes the near-surface region within a quasiparticle scattering length. In this regard, it is usually desirable to obtain statistics of the results from different spots on the sample. It is thus possible to check the homogeneity of the investigated property or, on the other hand, the presence of impurities, dead layers, damaged or strongly oxidized regions. When using the standard needle-anvil technique, a tip is brought in contact with a sample by a micrometer screw or piezoelectric element\cite{park06} but, most importantly, it can usually move only along the vertical ($z$) direction. So, every time a different region of the sample needs to be probed, a complete thermal cycle and positioning of the tip have to be repeated, adding a considerable amount of time (and money in the case of liquid He cooling). The addition of at least one degree of freedom for the movement of a tip, say, along the lateral direction $x$, would help to greatly reduce the time necessary to collect statistics of the spectra from different places of the sample. Moreover, besides helping also to better \ic tune'' the mechanical characteristics of a single point-contact, the in-plane motion of the tip would be very helpful also in case of double junctions in an interferometer configuration mentioned earlier, especially when the contacts are formed on different facets of a single crystal.\cite{brawner94} In that kind of experiments, the critical current of the two junctions, which also depends on the contact area, has to be as equal as possible in magnitude, in order to maximize the modulation of the critical current in the interferometer as a function of  magnetic flux through the interferometer itself,\cite{golubov13} and the piezoelectric nano-positioners provide that tunability.

In this paper we present the design and performance test of a PCS probe where the tip motion can occur both along the $z$ and $x$ axes. The head of the probe presented here can be unmounted from the probe and thus be used with many different cryostats, but it has been specifically designed for Quantum Design PPMS$^{\circledR}$ Dynacool$^{\textrm{TM}}$ system. The head can also be enclosed in a sealed cap in order to protect air-sensitive samples during the transfer into a cryostat. Furthermore, it has also been designed for hosting a coaxial cable for microwave irradiation of the sample and a small Helmholtz coil for applying low transverse magnetic fields. In this way, the probe is also suitable for studying point-contact Josephson junctions or performing phase-sensitive experiments. The probe has been tested by conducting PCS on superconducting Nb thin films and the conductance spectra were recorded as a function of temperature and applied magnetic field.

\begin{figure}[btp]
\begin{center}
\includegraphics[keepaspectratio, width=0.9 \columnwidth]{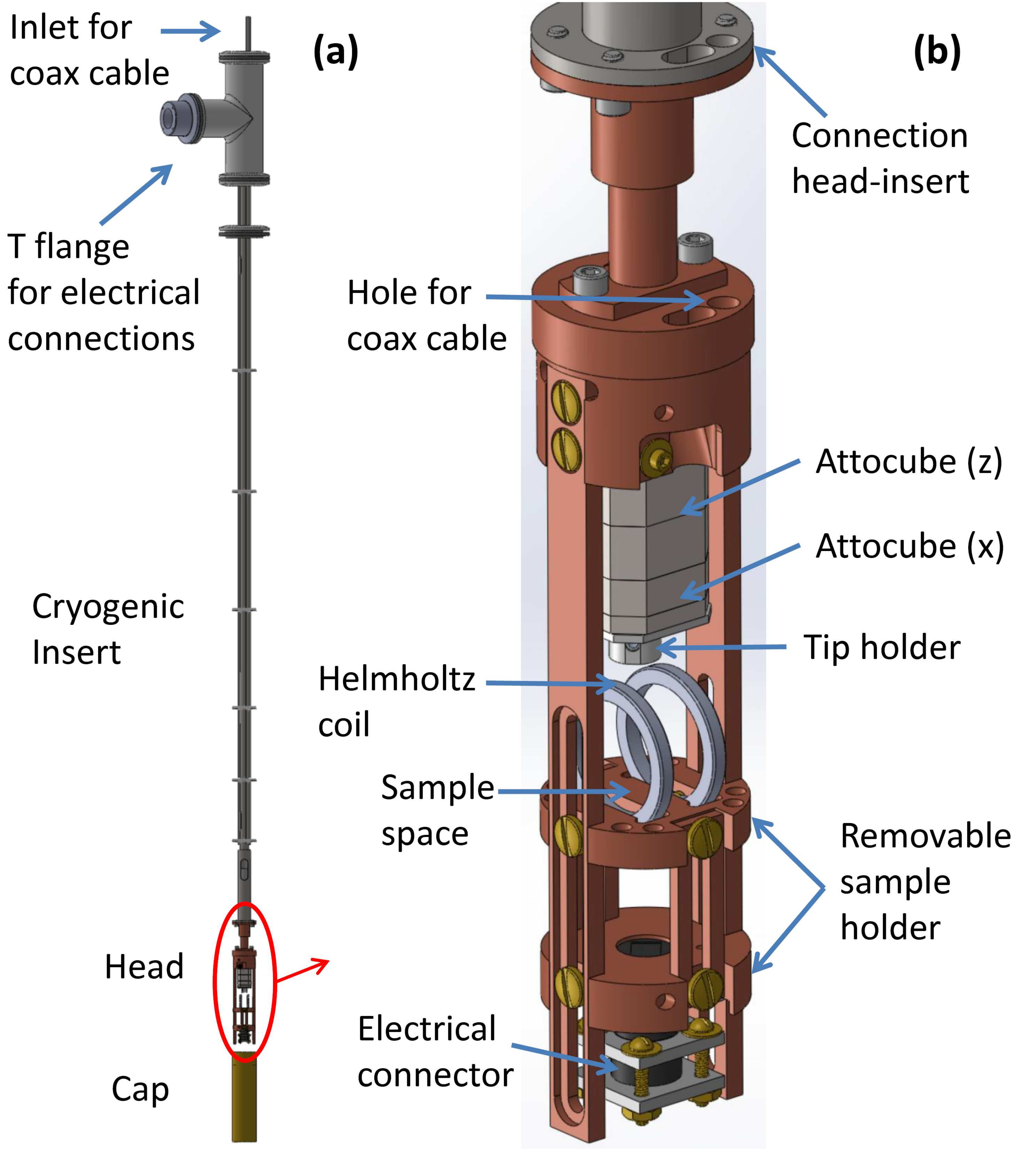}
\end{center}
\vspace{-5pt}
\caption{(a) Scheme of the whole probe whose total length is about 110 cm. This probe has been specifically designed for being used in a Quantum Design PPMS$^{\circledR}$ Dynacool$^{\textrm{TM}}$ system. (b) Zoomed view of the head of the probe. The head can be removed from the probe and used with different cryostats. Two Attocube piezo nano-positioners allow two degrees of freedom for the tip motion. The sample holder can be separated from the head and a sample is mounted while the holder is kept in the vertical position and/or inside a glove box/bag. The head is also arranged for hosting a Helmholtz coil and a coaxial cable plus, if necessary, additional wiring.}\label{probe}
\end{figure}

\section{\label{sec:second}Design}
\subsection{\label{sec:probe} PCS probe}

Figure \ref{probe}(a) shows a schematic of the whole probe, whose length is about 110 cm. The maximum diameter that goes into the vacuum chamber of the cryostat is that of the sealing cap, that is, 2.65 cm. Most of the wiring enters from a vacuum-sealed connector mounted on the side opening of a T-type flange but an inlet has also been added at the top to allow for additional wiring for, e.g., a semi-rigid coaxial cable (whose diameter can be up to 3.9 mm) for microwave irradiation of point-contact Josephson junctions.\cite{burmistrova15} The measurement and thermometer wires are a cryogenic ribbon cable Cryoloom$^{\circledR}$  from CMR-Direct. The cable has been wound along the cryogenic insert in order to minimize the heat carried from the laboratory ambient temperature to the cryogenic sample space. Near the bottom of the probe, the ribbon cable is fixed to an electrical connector in order to allow the head to be separated and attached on different inserts suitable for other cryostats.

Figure \ref{probe}(b) shows a schematic of the probe head. The tip holder is fixed on a stack of two linear nano-positioners from Attocube Systems AG. The top one is model ANPz51/LT and moves the tip along the vertical ($z$) direction with maximum travel distance of 2.5 mm. On that, an ANPx51/LT positioner is fixed for the movement of the tip along the lateral ($x$) direction (up to 3 mm). Both positioners have a footprint of 15 mm x 15 mm and are made of non magnetic titanium. In fact, all the probe materials are non magnetic to allow experiments in the presence of applied magnetic fields. The tip holder is made of an aluminum cylinder which is fixed by means of a screw to a Teflon support, in order to electrically decouple the tip from the metallic positioner plate. The tip is kept in place by a second screw which is connected to the electrical wires that create the connection with the instruments. The sample holder consists of two round copper pucks held together by two flat supports. A sample is placed on the top one and its electrical leads are connected to a series of pins fixed on the edge of the puck. Up to two samples can be mounted at the same time. A Cernox thermometer (Lake Shore Cryotronics, Inc.) is mounted in a cavity right below the sample. The lower puck hosts a connector that allows removing the sample holder from the head. In this way the sample can be comfortably mounted by keeping the holder in the upright position. For point-contact Josephson junction or phase-sensitive measurements, a similar sample holder is constructed such that it can also host a Helmholtz coil in two cavities built into the copper puck, allowing a small transverse magnetic field to be applied. The Helmoltz coil is a convenient setup for obtaining a uniform magnetic field in its inner region.\cite{ramsden} In a Helmoltz coil the distance between the two sub-coils equals their radius and the one shown in Fig. \ref{probe}(b) has been designed considering a radius of each sub-coil of about 8 mm. In this way, by using a superconducting wire it would be possible to achieve a field of more than 30 Gauss with, for instance, 150 turns and a current of 180 mA. \cite{ramsden} When dealing with air-sensitive samples, the probe head can be kept in a glove box/bag and, after a sample has been placed on the holder, the sample can be protected during the quick transfer to the cryostat by putting a sealable brass cap over the head.

The design presented here, which combines in a single apparatus the lateral scanning capability along with the possibility of applying transverse magnetic fields and microwave irradiation, makes it suitable for at least three different kinds of experiment: site-dependent PCS, Josephson-junction and phase-sensitive measurements. Another noticeable difference from previous works like, for example, that of Groll \emph{et al.} \cite{groll15} is that in that case they have three degrees of freedom ($x$, $y$, and $z$) and a full \ic mapping" capability. This means that the piezoelements employed are like those used in scanning probe microscopes, which have a precise closed-loop control of the stroke of the piezo elements. Those elements have bigger size and higher cost. For our purposes, a simpler and cheaper solution is preferable where only one additional, open-loop, lateral degree of freedom is sufficient. Moreover, the size of our apparatus and in particular the diameter of the head is smaller than in Ref. 28. Indeed, while in Ref. 28 the maximum sample diameter is 25 mm, in our case the whole head has a diameter of 26.5 mm. This is because our rig has been specifically designed for the Quantum Design PPMS$^{\circledR}$ Dynacool$^{\textrm{TM}}$ system, whose vacuum chamber has an inner diameter of about 28 mm, posing stricter constraints on the use of the limited space available.

\subsection{\label{sec:ElectronicsControl} Electronics and control software}

\begin{figure}[btp]
\begin{center}
\includegraphics[keepaspectratio, width=0.9 \columnwidth]{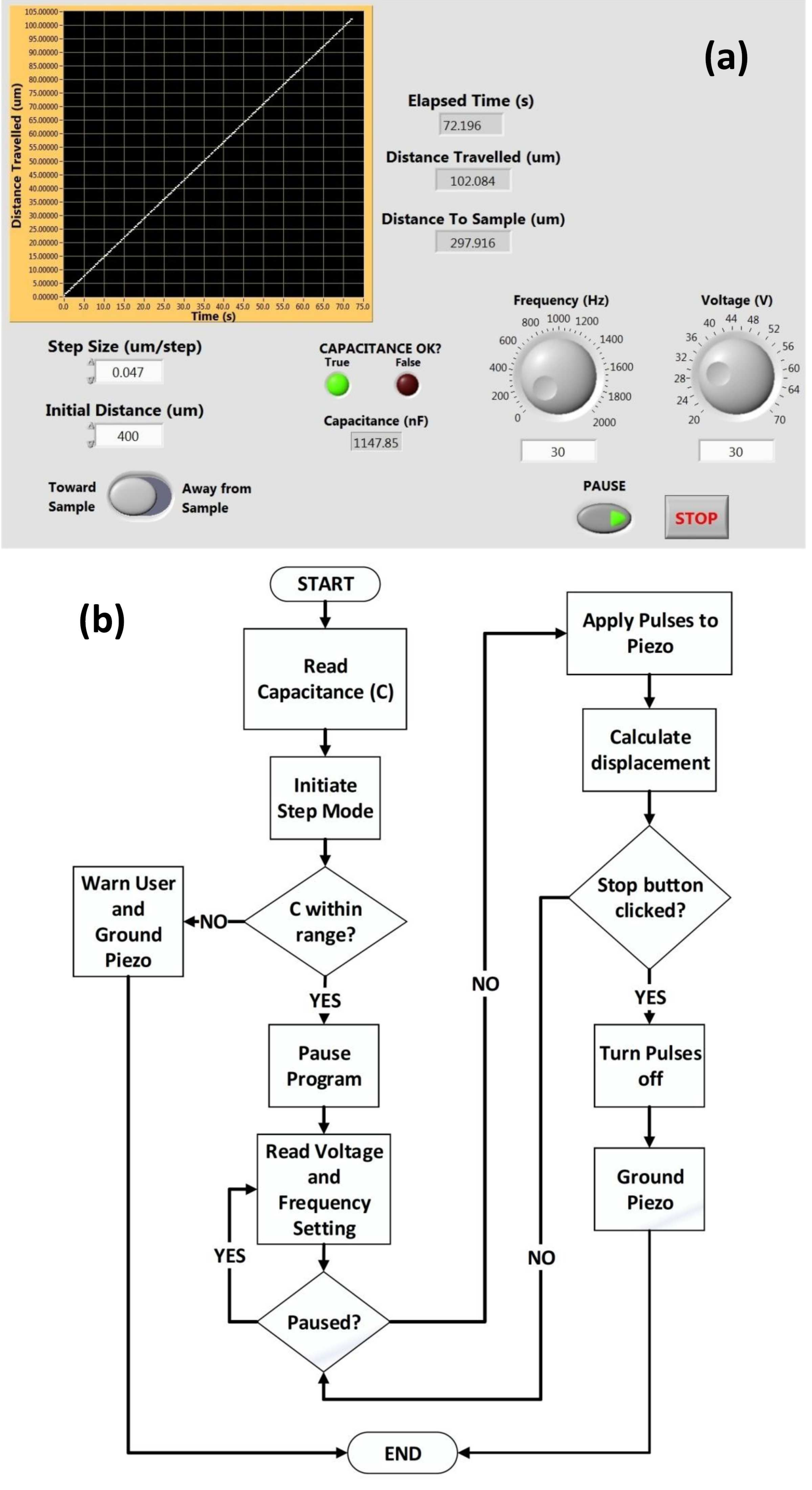}
\end{center}
\vspace{-5pt}
\caption{(a) LabVIEW front panel for the control of the $z$ nano-positioner. A similar panel exists for the $x$ positioner as well. The user can set different parameters (frequency, voltage, distance, etc.) in order to control and tune the approach of the tip towards or away from the sample ($z$) or for moving the tip transversely to probe different areas of the sample ($x$). (b) Flow chart for the positioner control procedure. The measured positioner capacitance (C) is used to detect short circuits or other potential electrical problems with the positioner.}\label{electronics}
\end{figure}

The hardware and software used to realize the piezoelectric open-loop control system consists of an Attocube ANC300 piezo controller and a pair of control programs (one for each degree of freedom) implemented in the LabVIEW System Design Software, respectively. In this configuration, there is continuous serial communication between the measurement computer and the ANC300 controller via a USB cable. The purpose of the ANC300 piezo controller is to provide a train of voltage pulses which drive the piezoelectric motion via discrete displacements (steps).  For a 30 V peak-to-peak pulse train, the step size at 1.74 K was determined to be, approximately, 0.047 $\mu$m. The front panel for the $z$ positioner control program is shown in Fig. \ref{electronics}(a). Here the user can make adjustments to the controller settings when attempting to form a metallic junction. A similar front panel exists for the $x$ positioner control program (not shown here). Because temperature-dependent data can be used to determine the stability of a point-contact junction formed, the sample temperature is read with the use of a Lake Shore Cernox thermometer and the status of the cryostat is monitored via a local area network interface between the measurement computer and the Dynacool system. The flow chart for a basic positioner control program is shown in Fig. \ref{electronics}(b). It is identical for both $x$ and $z$ positioner control programs.

\begin{figure}[btph]
\begin{center}
\includegraphics[keepaspectratio, width=0.9 \columnwidth]{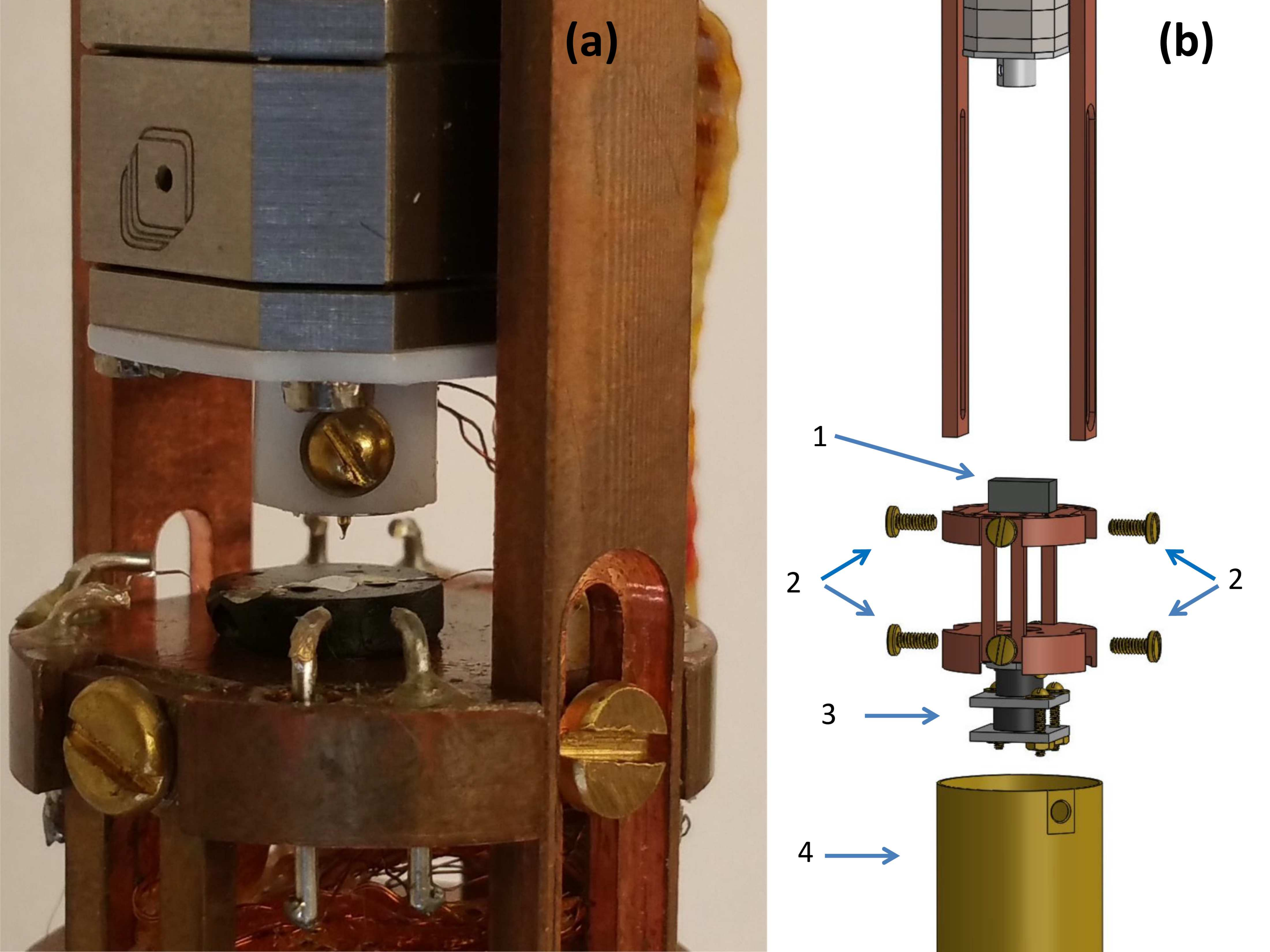}
\end{center}
\vspace{-5pt}
\caption{(a) Side view of the head of the probe. The tip holder mounts an Au tip. The sample holder is now supporting a test sample which is also electrically connected to the pins. (b) Scheme showing how to remove the sample holder. 1: Sample. 2: Screws for attaching and removing the sample holder. 3: Detachable electrical connector (head to sample holder). 4: Brass cap that can be used to protect air-sensitive samples.}\label{pic}
\end{figure}

The formation of a junction between an electrochemically sharpened metallic tip and a metallic sample surface is done as follows. With the sample and tip mounted on the head, the probe is inserted into the Dynacool cryostat and, once the target temperature ($\approx 2$ K, typically) is reached, the $z$ positioner is slowly stepped in the direction of the sample surface while the junction resistance (obtained via a standard four probe lock-in technique) is continuously monitored on the   front panel of a LabVIEW conductance measurement program. Once a suitable junction resistance is achieved, the displacements are terminated and the intended experiment can be performed. Multiple junctions can be tested by the use of the $x$ positioner. For this, the tip is raised to a safe distance above the sample surface using the $z$ positioner and the $x$ positioner is stepped in the desired direction. Once the required $x$ displacement is achieved, the $z$ positioner is used to bring the tip into contact with the sample to form a new junction. Figure \ref{pic}(a) reports a close view of the tip and the sample region, when the gold tip is raised at a large distance from the sample. Panel (b) shows instead a schematic drawing on how to exchange the sample. The sample holder is detached from the head by removing four screws and an electrical connector. The holder can then be placed on a lab bench in order to comfortably mount the sample to be measured. If needed, a brass cap can be screwed to the head in order to protect air-sensitive samples.

\section{\label{sec:4}Performance and Discussion: Demonstration with measurements on Nb}

\begin{figure}[btph]
\begin{center}
\includegraphics[keepaspectratio, width=0.9 \columnwidth]{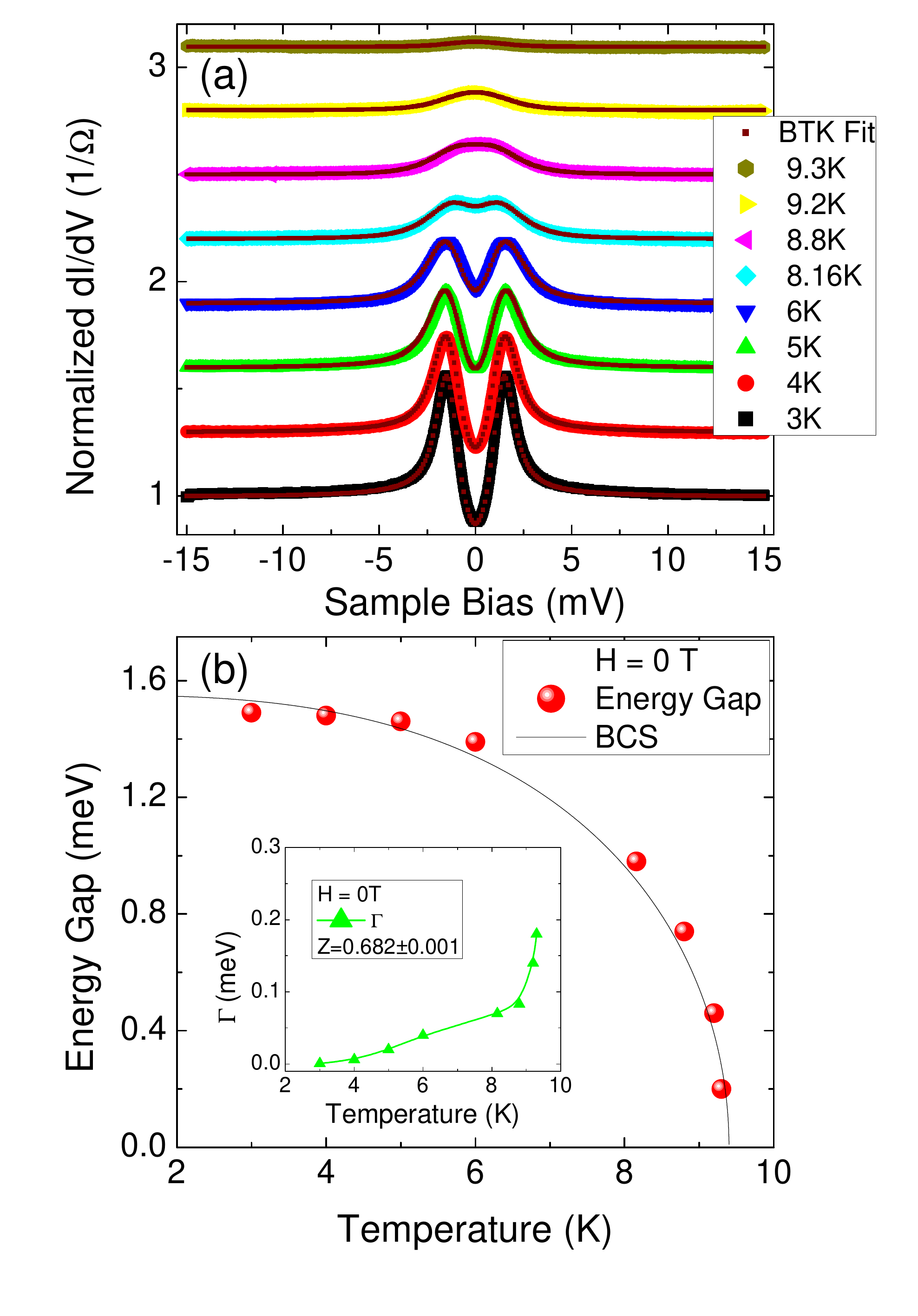}
\end{center}
\vspace{-5pt}
\caption{(a) Temperature dependence of the normalized conductance (colored symbols, shifted vertically for clarity) for an Au/Nb point contact. The brown squares are the BTK fits to the experimental spectra. (b) Temperature dependence of the energy gap $\Delta$ as determined by the BTK fits presented in (a). The solid line represents a BCS-like behavior of $\Delta(T)$. The inset shows the $\Gamma$ parameter used for the BTK fits as a function of temperature. It increases approximately linearly up until near the $T_c$ of Nb.}\label{Tdep}
\end{figure}

In order to test the performance of the probe, we conducted point-contact Andreev reflection spectroscopy (PCARS) measurements on Nb thin films. The Nb films were deposited onto $c$-plane sapphire substrates using DC magnetron sputtering. The film thickness was 2,000 $\AA$. The residual resistivity ratio was about 100 and the critical temperature ($T_c$) was 9.25 K. They were then mounted on the removable sample holder and electrically connected to it, as shown in Figure \ref{pic}(b). As the normal metal for the point contact, an electrochemically sharpened Au tip was employed.\cite{narasiwodeyar15} In particular, the tips used here are made by a two-step electrochemical etching process: the first step is with HCl mixed with glycerol and the second step is with pure HCl. An Au wire of diameter 0.5 mm is immersed into the mixture solution. A DC offset of 6 V and AC pulse of 1 kHz, 3 V peak to peak amplitude is applied between the Au wire and Pt counter-electrode. The stopping current is set to 54 mA. During this process the meniscus formed around the Au wire causes the Au wire to be etched into a long and thin tip. The second part of the etching process is used to eliminate the rougher surface caused by the meniscus from the first etching process. The tip is re-immersed into a pure HCl solution and a constant voltage of 6 V is supplied for 100 ms. Since the pure HCl does not form a strong meniscus, the effects of the meniscus are negligible and this process only improves the surface quality of the tip. The tip is then removed and cleaned with unheated de-ionized water and methanol. Further details on the tip fabrication can be found in Ref. 29. After attaching the sample holder to the head, the probe was inserted inside the Dynacool and cooled down to 3 K. As reported in section \ref{sec:second}, the junction was formed at low temperature by progressively moving the tip towards the sample by means of the $z$ nano-positioner. In this way it was possible to form ballistic contacts and observe the Andreev reflection phenomenon, as shown in Fig. \ref{Tdep}(a), which represents the normalized differential conductance. The shape of the curve, in particular, the absence of dips outside the gap\cite{daghero10}, is also a signature of the good spectroscopic nature of the junction. The temperature was then increased and the differential conductance was recorded at different temperatures up to 9.3 K. The two characteristic maxima in the conductance gradually decrease, merging into a single peak which eventually disappears at the critical temperature. The temperature dependence of normalized conductance is shown in Fig. \ref{Tdep}(a). The spectra were then fitted to a generalized\cite{plecenik94, dynes78} Blonder-Tinkham-Klapwijk (BTK) model.\cite{BTK} The model contains, besides the temperature $T$, three free parameters called $\Delta$, $\Gamma$, and $Z$. $\Delta$ is the amplitude of the order parameter. $\Gamma$ is the broadening parameter and accounts for the finite lifetime of the quasi-particles and for inelastic scattering processes at the normal metal/superconductor (N/S) interface. $Z$ is the dimensionless barrier strength and denotes the presence of a potential barrier (like an oxide layer) at the N/S interface.

The brown squares in Figure \ref{Tdep}(a) represent the BTK fits to the normalized conductance curves. It is clear that the fitting curves overlap with the experimental data very well at all temperatures, indicating good quality of the fit. Red symbols in Fig. \ref{Tdep}(b) represent the $\Delta$ parameter as obtained by the fits shown in \ref{Tdep}(a). $\Delta$ gradually decreases with increasing temperature following a BCS-like behavior (solid line), from which a zero temperature value of the gap can be inferred, $\Delta(0) = 1.55$ meV, in good agreement with other results reported in literature.\cite{novotny75, kittelbook} As expected, $Z$ is constant with temperature while $\Gamma$  increases slightly with temperature.

\begin{figure}[btph]
\begin{center}
\includegraphics[keepaspectratio, width=0.9 \columnwidth]{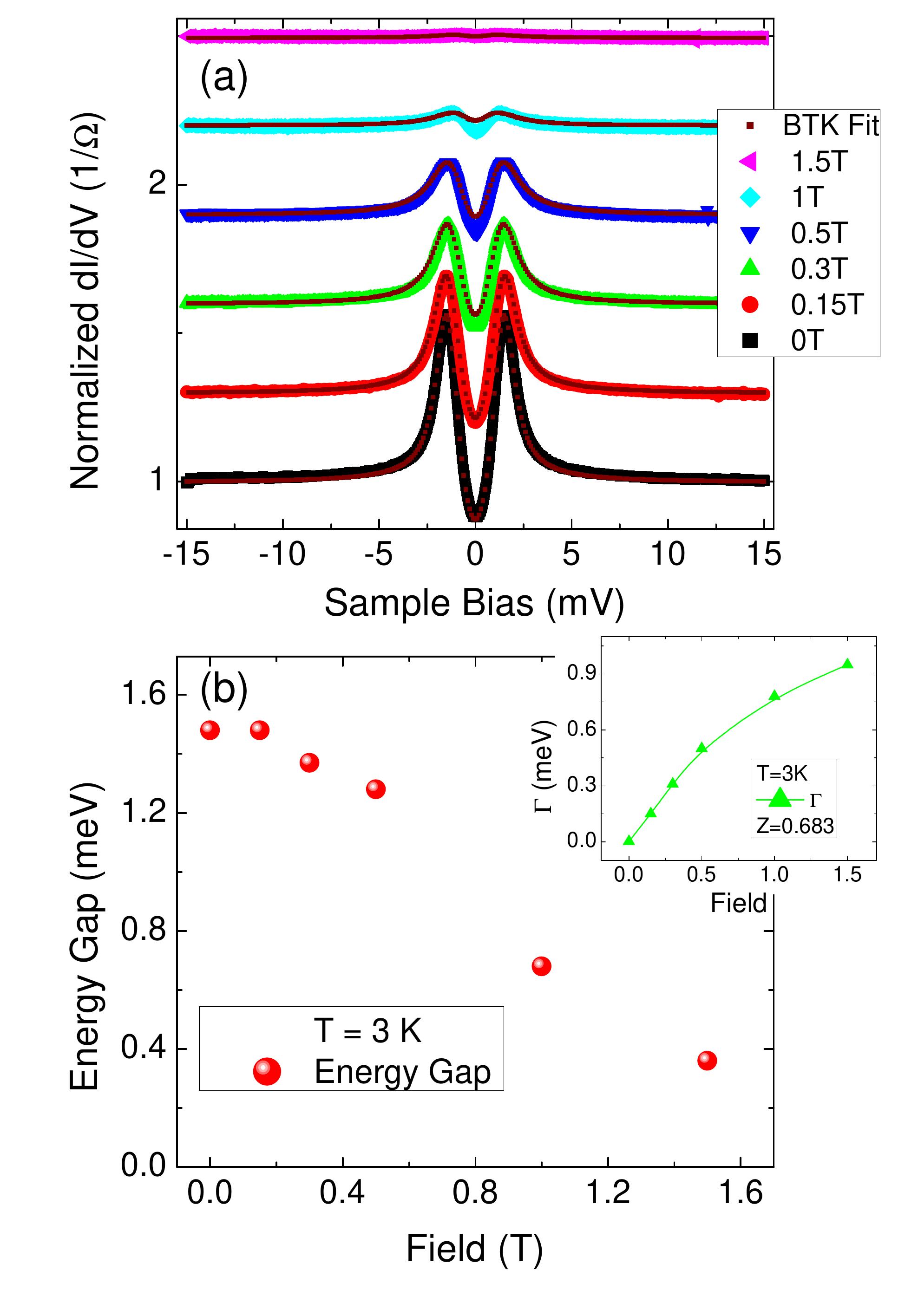}
\end{center}
\vspace{-5pt}
\caption{(a) Magnetic field dependence of the normalized conductance (colored symbols, shifted vertically for clarity) for an Au/Nb point contact. The brown squares are the BTK fits to the experimental spectra. (b) Magnetic field dependence of the energy gap $\Delta$ as determined by the BTK fits presented in (a). The inset shows the $\Gamma$ parameter used for the BTK fits as a function of magnetic field. It increases approximately linearly in the low field region. }\label{Bdep}
\end{figure}

PCARS measurements were also conducted in the presence of a longitudinal (with respect to the probe axis) static magnetic field to check whether the probe works well also under a magnetic field. The experiment was successful, as shown in Fig. \ref{Bdep}(a), where the magnetic field dependence of the normalized conductance at 3 K is reported (colored symbols). The effect of an increasing magnetic field on the Andreev spectra is apparently similar to that of the temperature, i.e., the two peaks move to lower bias voltages with concomitant decrease in their heights and then, approximately at the critical field of the Nb thin film, the conductance curve flattens out and becomes similar to the curve taken above $T_c$. The brown squares in Fig. \ref{Bdep}(a) represent the BTK fits to the experimental spectra. Again, the BTK model reproduces the experimental data very well. As it has been shown by Naidyuk \emph{et al.},\cite{naidyuk96} at finite temperatures the pair breaking effect of the magnetic field can be, in the first order approximation, modeled by an extra contribution $\Gamma_f(B)$ to the intrinsic field independent $\Gamma$ parameter.\cite{daghero10} As for the spectra shown in Fig. \ref{Bdep}(a), the broadening parameter as a function of magnetic field is shown in the inset of Fig. \ref{Bdep}(b). It increases up to about 1 meV with increasing magnetic field. The main panel reports the behavior of the gap as a function of magnetic field (red symbols). The energy gap decreases with increasing magnetic field and becomes zero for fields higher than 1.5 T. This value is considerably larger than the upper critical field reported in the literature for bulk Nb.\cite{butler80} This can be explained by the large demagnetization effect when the magnetic field is perpendicular to the thin film plane (the demagnetization factor $\sim$ 1), as it is the case here.

\begin{figure}[btph]
\begin{center}
\includegraphics[keepaspectratio, width=0.9 \columnwidth]{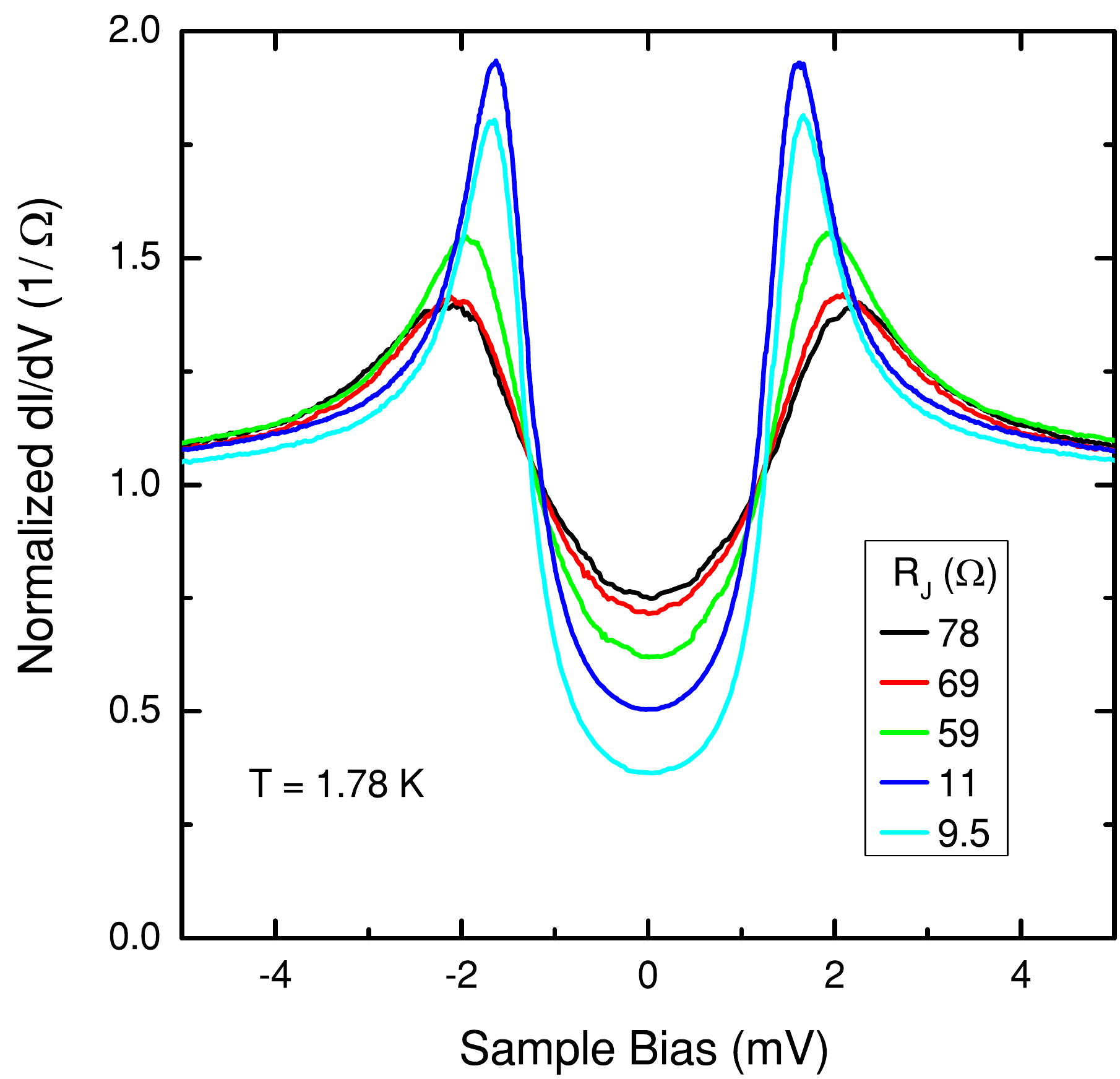}
\end{center}
\vspace{-5pt}
\caption{Normalized conductance curves obtained at five different sample sites thanks to the lateral scanning capability of the rig. The resistance of each of the five Au/Nb point-contact junctions is indicated in the figure.}\label{Statistics}
\end{figure}

Figure \ref{Statistics} shows an example of site-dependent PCARS measurements. The conductance curves were obtained from different point-contacts formed by moving the tip along the $x$-direction by 100 $\mu$m each time starting from the initial one of junction resistance $R_J$ = 78 $\Omega$. They all look BTK-like with $Z = 1 \sim 2$, indicating the existence of surface oxides of sizable thickness. The conductance shape varies systematically in correlation with $R_J$, which was adjustable with the tip-sample pressure. The junction resistance is most probably inversely correlated with the size of the contact area that increases as the tip is pressed more, and vice versa. It can also be noticed that, as the resistance decreases, the conductance curves clearly become less smeared. This is due to the fact that the profile of the potential barrier at the junction interface becomes sharper with increasing the tip-sample pressure, making the conductance curves cleaner and less blurred. \cite{blonder83} Since the systematic evolution of the conductance shape can be explained in terms of different degrees of smearing governed mostly by the tip-sample pressure, we can conclude that the sample itself is quite uniform over the probed region. This measurement clearly demonstrates that indeed the added degree of freedom along the lateral direction works well enabling to obtain statistics over large areas.

\section{\label{sec:5}Summary}
In summary, we have reported the design, realization, and test performance of a cryogenic probe for PCS in the standard needle-anvil configuration. By employing two suitable piezo nano-positioners from Attocube Systems AG, the probe provides two degrees of freedom for the tip motion, along $x$ and $z$, enabling to probe different areas of a sample \textit{in situ} or to form double point contacts on different facets of a superconducting single crystal for the purpose of Josephson interferometry. The probe is specifically designed for a Quantum Design PPMS$^{\circledR}$ Dynacool$^{\textrm{TM}}$ system but the head can be separated and used in different cryostats or placed in a glove box/bag for the purpose of protecting an air-sensitive sample until transferring into a cryostat. Moreover, it has also been arranged for inserting a coaxial cable for microwave irradiation (e.g. for the study of Josephson junctions) and for hosting a Helmholtz coil for applying a small transverse magnetic field. This can be useful, for instance, for performing phase-sensitive experiments that make use of single or double point contact(s).\cite{brawner94,golubov13} The performance of the probe has been successfully tested by conducting PCARS experiments on Nb thin films over large areas as a function of temperature and longitudinal static magnetic field, and by analyzing the obtained conductance spectra using the generalized BTK model.\\

\section{\label{Acknowledgments}Acknowledgments}
We thank E. P. Northen and J. Brownfield for their work in the machine shop.
This work was supported by the Center for Emergent Superconductivity, an Energy Frontier Research Center funded by the U.S. DOE, Office of Science, Office of Basic Energy Sciences under Award No. DE-AC0298CH1088. W.K.P., C.O.A., and P.S. were supported by the U.S. NSF DMR under Award No. 12-06766. M.T. acknowledges the U.S.-Italy Fulbright Commission, Core Fulbright Visiting Scholar Program, during which the apparatus was designed.

\bibliography{Paper_AC-PCS_RSI_arXiv.bbl}

\end{document}